\documentclass[twocolumn]{aastex63}

\received{}
\revised{}
\accepted{\today}
\submitjournal{ApJ}

\shorttitle{Kraken multiframe deconvolution at LBT}
\shortauthors{Hope et al.}

\usepackage{lineno}
\begin{document}

\title{Post-AO high-resolution imaging using the Kraken multi-frame blind deconvolution algorithm}

\correspondingauthor{Douglas A. Hope}
\email{Douglas.Hope@gtri.gatech.edu}

\author{Douglas A. Hope}
\affiliation{Georgia State University, 25 Park Place, Atlanta, GA 30303, USA}

\author{Stuart M. Jefferies}
\affiliation{Georgia State University, 25 Park Place, Atlanta, GA 30303, USA}
\affiliation{University of Hawaii, 34 Ohia Ku Street, Pukalani, HI 96768, USA}


\author{Gianluca Li Causi}
\affiliation{INAF-IAPS, Via del Fosso del Cavaliere, 100, 00133 Roma, Italy}
\affiliation{INAF-OAR, Via Frascati, 33, 00040 Monte Porzio Catone (RM), Italy}
\affiliation{INAF-ADONI, Adaptive Optics National Laboratory, Italy}

\author{Marco Landoni}
\affiliation{INAF - Osservatorio Astronomico di Cagliari. Via della Scienza - Selargius (CA) - Italy}
\affiliation{INAF - Osservatorio Astronomico di Brera, via E Bianchi 46 Merate (LC) - Italy}



\author{Marco Stangalini}
\affiliation{ASI Italian Space Agency,Via del Politecnico snc, 00133, Rome, Italy}
\affiliation{INAF-OAR, Via Frascati, 33, 00040 Monte Porzio Catone (RM), Italy}

\author{Fernando Pedichini}
\affiliation{INAF-OAR, Via Frascati, 33, 00040 Monte Porzio Catone (RM), Italy}
\affiliation{INAF-ADONI, Adaptive Optics National Laboratory, Italy}

\author{Simone Antoniucci}
\affiliation{INAF-OAR, Via Frascati, 33, 00040 Monte Porzio Catone (RM), Italy}
\affiliation{INAF-ADONI, Adaptive Optics National Laboratory, Italy}




\begin{abstract}

In the context of extreme adaptive optics (ExAO) for large telescopes, we present the Kraken multi-frame blind deconvolution (MFBD) algorithm for processing high-cadence acquisitions, capable to provide a diffraction-limited estimation of the source brightness distribution. This is achieved by a data modeling of each frame in the sequence driven by the estimation of the instantaneous wavefront at the entrance pupil. Under suitable physical contraints, numerical convergence is guaranteed by an iteration scheme starting from a Compact MFBD (CMFBD) which provides a very robust initial guess which only employs a few frames. We describe the mathematics behind the process and report the high-resolution reconstruction of the spectroscopic binary $\alpha~And$ ($16.3~mas$ separation) acquired with the precursor of SHARK-VIS, the upcoming high-contrast camera in the visible for the Large Binocular Telescope.

\end{abstract}



\section{Introduction} \label{sec:intro}

Turbulence in the Earth's atmosphere unavoidably degrades the sharpness of astronomical images obtained with ground-based telescopes.  The turbulence distorts the planar wavefront coming from the astronomical point sources and prevents the formation of the ideal Airy diffraction pattern at the telescope focal plane. 
Fortunately, by using Adaptive Optics (AO), an advanced instrumentation technique developed over the last few decades, we can partially correct these distortions in real-time. The Large Binocular Telescope (LBT) currently hosts one of the most advanced extreme AO systems (ExAO) built to date, called FLAO for First Light Adaptive Optics \citep{Esposito_2010, Pinna_2015}. The LBT's ExAO corrects the first 550 modes of the wavefront thanks to its Pyramid Wavefront Sensor \citep{Ragazzoni_1996} at a cadence of 1kHz with 3ms of lag time that, in the visible bands (550-850nm), provides a sharp point spread function (PSF) with a core width very close to diffraction limit (rms wavefront error of about 90-100nm on stars brighter than mag 6).

\begin{figure*}[ht]
    \centering
    \includegraphics[width=18cm]{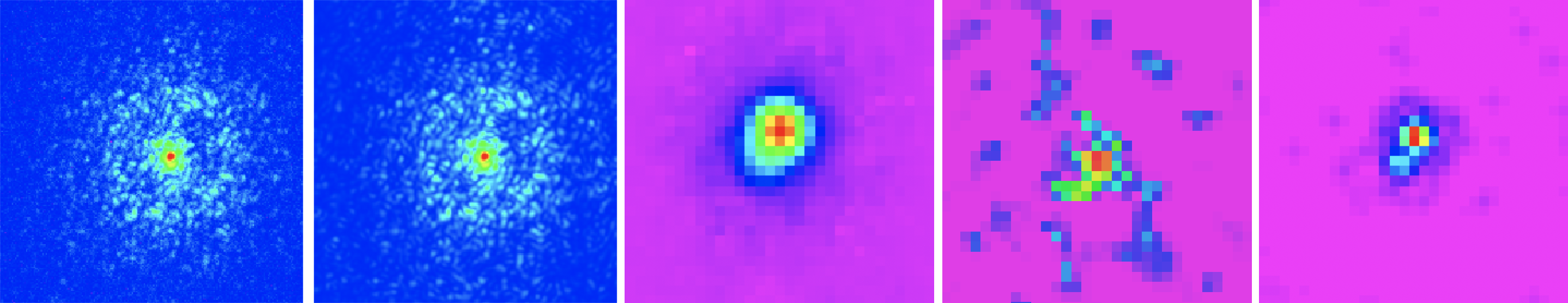}
    \caption{Data modeling and object estimation in MFBD. Panels from left to right: one of the selected frames (i) and its Kraken model (ii); SFI reconstruction from \citet{Mattioli_2019} (iii); CMFBD initial estimate based on the best 2 frames (iv); full Kraken MFBD reconstruction (v). 
}
    \label{fig:kraken_vs_sfi}
\end{figure*}

Performance of ExAO correction is commonly measured by the ``Strehl ratio" \citep{Roggemann-1996}. For the best frames we have acquired at the LBT in the visible band the Strehl ratio is around 0.3.
This highlights the fact that, unfortunately, even ExAO correction is never perfect, and the resultant post-AO imagery does not reach the diffraction limit: a significant fraction of the starlight remains in many ``speckles" outside the sharp PSF core, as depicted in the first panel of figure \ref{fig:kraken_vs_sfi}. Moreover, the morphology of these speckles rapidly evolves with a typical timescale of a few ms (see \citet{Stangalini_2016}). 

In practice, such varying speckle patterns prevent us from reaching the theoretical angular resolution of the telescope, thus also reducing the contrast. They also block us from resolving closely spaced sources, especially when dealing with faint sources close to bright objects.
Digital post-processing of the images acquired after the adaptive optics can effectively help to recover the full spatial resolution of the telescope, so long as the speckles' evolution is measured with a sufficiently fast temporal cadence in order to freeze the PSF in each frame.

As a part of the SHARK-VIS project \citep{Pedichini_2016, Mattioli_2018}, the high constrast visible imager at the Large Binocular Telescope (LBT), we are investigating new post-processing techniques, e.g. Speckle-Free Imaging and Speckle-Free Angular Differential Imaging (SFI and SFADI and, \citet{Li_Causi_2017, Mattioli_2019}), with the aim of both achieving diffraction-limited resolution performance and noise-limited high contrast imaging at sub-arcsecond separation.\\
In this context, we present here the first astronomical results obtained with a novel implementation of the Multi-Frame Blind Deconvolution (MFBD) algorithm, called Kraken MFBD \citep{Hope2016}. 
The Kraken MFBD algorithm was initially developed for high-resolution, high-contrast imaging of Earth-orbiting artificial satellites, which, like astronomical sources, suffers from significant image degradation due to atmospheric turbulence.

\section{The Kraken MFBD algorithm}
MFBD is a known mathematical technique \citep{Jefferies1993, Schulz1993} able to estimate both the object and the PSF evolution directly from the recorded images, under the only requirement that the object is stationary during the observations and the changing PSF demonstrates significant diversity. Although MFBD proved to surpass the performance of other speckle imaging methods \citep{Matson2008}, and is nowadays routinely used in solar observations \citep{2005SoPh..228..191V}, its use has not gained traction in the nigthtime astronomy community. This lack of traction is due to the high computational cost of the MFBD algorithm, difficulties associated with processing many images, and uncertainty in the reconstructions’ photometric quality. Research to overcome these obstacles has been limited, but our recent studies have provided several algorithmic improvements, that allow the current Kraken software to address these problems and provide state-of-the-art reconstructions.

 We have discovered that restoring images acquired through strong atmospheric turbulence is often best achieved using a variety of restoration techniques in combination.  Our software is built on this premise and provides access to several restoration approaches and the combinations of these approaches, which include: Basic MFBD, Compact MFBD \citet{Hope2011, Hope2019}, Myopic MFBD with WFS data \citet{Conan:2000}, Aperture diversity \citet{Hope2016}, Wavelength diversity \citet{Ingleby:05}, and Phase diversity \citet{Paxman:92}. 
 
Probably the largest shortcoming of MFBD algorithms is their sensitivity to entrapment in local minima; especially, when using many frames in the restoration process. CMFBD provides a means to produce good initial estimates for the object and PSFs whilst using a greatly reduced number of variables, reducing the susceptibility to entrapment in local minima.   We have found that a multi-step restoration process in which the estimates for the object and PSFs are refined before committing to a full MFBD restoration provides the best performance. We note that myopic MFBD and the diversity imaging techniques supported in Kraken, provide a constraint on the atmospheric wavefront, and thus the PSFs for the images. These constraints thus provide strong leverage for separating the object from the PSFs by MFBD.

One development is Compact MFBD (CMFBD: \citet{Hope2011, Hope2019}), a means to acquire high-quality initial estimates of the object and PSFs for the MFBD, by using a relatively small number of variables and prior information on the nature of the PSFs. Another development is modeling the PSF by estimating the real and imaginary parts of a band-limited function whose cut-off frequency is commensurate with the telescope’s cut-off frequency \citep{Miura2003}, a model able to accommodate any wavefront amplitude variation faster than the atmospheric coherence time, as well as the errors incurred from using a Fourier Optics model of the PSF for long exposure times.

In practice, the basic Kraken MFBD algorithm estimates the blur-free object and the PSFs for a set of data frames by minimizing the error metric
\begin{equation}
    \epsilon = \Sigma_l \Sigma_{x,y} \left[(g_l(x,y)-f(x,y)\odot h_l(x,y))/\sigma_l(x,y) \right]^2
    \label{eqn:mfbd}
\end{equation}
Here $g(x,y)$ is an observed image, $f(x,y)$ is the object, $h(x,y)$ is the point spread function, $\sigma(x,y)$ is a weighting term that depends on the noise in the observed image, $\odot$ denotes convolution, and $l$ is the number of frames.

The object is modeled via the reparameterization $f(x,y)=\phi^2(x,y)$ and the PSF is modeled via the wavefront amplitudes, $A(u,v)$, and phases, $\psi(u,v)$. That is, $h(x,y) = FT^{-1}\left[P(u,v) \star P(u,v)\right]$ where $P(u,v) = A(u,v) \exp{(-i\psi(u,v))}$ and $\star$ denotes correlation. 

The quality of an MFBD restoration is expected to improve as the number of frames ($l$) used in the restoration increases (because there is more information available). However, as the number of frames increases, so does the number of variables required to model all the data. Eventually, the estimated variables' quality deteriorates due to the large dimensionality of the parameter hyperspace that leads to inevitable entrapment in local minima during the optimization. These local minima can be far from the global minimum. To reduce the probability of entrapment in a local minimum, we use CMFBD to provide high-quality estimates for the object and PSFs for all the data frames. These estimates are then used in a traditional MFBD restoration. That is, we use CMFBD to provide estimates of $f(x)$ and $h_l(x)$ which are then iteratively updated using equation (\ref{eqn:mfbd}) and a conjugate gradients-based optimization algorithm.  

The CMFBD algorithm is a two-step process. In the following we use the Fourier transform pair identities $[g(x,y);G(u,v)],[f(x,y); F(u,v)]$ and $[h(x,y);H(u,v)]$.

\subsection{CMFBD: Step 1}
We control the number of variables that need to be estimated by restricting the restoration to a subset of the data that contains the selected best frames. Here the selection is done to provide the highest signal-to-noise signal across a large fraction of the object's Fourier spectrum, using the smallest number of frames. The downside to this approach, of course, is that it only uses a sub-set of the data, and therefore we are not capitalizing on all of the information that is available to us. However, we can use the non-selected data frames to provide additional constraints on the PSF estimates for the selected data frames without increasing the number of model parameters. That is, the parameters that describe the object are determined by the information available in the selected data frames. The parameters that describe the PSFs for the selected frames, on the other hand, are determined by the information in {\it all} of the images in the data set. We also invoke prior knowledge we have about the PSFs for the non-selected frames via two penalty functions. The first enforces the prior knowledge that PSFs are positive functions. The second enforces the prior knowledge that the PSFs should not depend on how they are estimated. We then minimize the error metric

\begin{eqnarray}
\epsilon & = & \Sigma_k \Sigma_{x,y} \left[(g_k(x,y)-f(x,y)\odot h_k(x,y))/\sigma_k(x,y) \right]^2 \nonumber \\
&+& \alpha \Sigma_k\Sigma_{j\ne k}\Sigma_{x,y} m_{j}(x,y)h_{j,k}^2(x,y)  \nonumber \\
&+& \gamma \Sigma_{u,v} \Sigma_k \Sigma_{k' \ne k}~ M_{k,k'}(u,v) \Sigma_j M_j^H(u,v) \nonumber \\
&|&\beta_{j,k}(u,v)H_k(u,v) - \beta_{j,k'}(u,v)H_{k'}(u,v)|^2
\label{eqn:step1}
\end{eqnarray}

Here the index $k$ is over the frame-selected images, the index $j$ is over the remaining images (the non-selected frames), and $k+j = l$ (the total number of frames). The penalty scalar factors $\alpha$ and $\gamma$ are set to 0.1 \citep{Hope2019}.
The second term in equation \ref{eqn:step1} is the non-control frame PSF non-negativity penalty function.
Here the estimate of the PSF for the $j^{th}$ non-control frame, estimated via the optical transfer function (OTF) for the $k^{th}$ control frame is given by
\begin{equation}
    h_{j,k}(x,y) = |FT^{-1}\left[ H_k(u,v) \beta_{j,k}(u,v)\right]|^2~,
\end{equation}
where
\begin{equation}
    \beta_{j,k}(u,v)=G_j^{obs}(u,v)/G_k^{obs}(u,v)
\end{equation}
and
$m_j(x,y)=1$ if $h_{j,k}(x,y) < 0$, and $m_j(x,y) =0$ otherwise.
The third term in equation (\ref{eqn:step1}) is a PSF consistency metric which enforces the prior knowledge that the PSF estimate for a non-control frame should not depend on which control frame is used to estimate it.
In this term $M_{k,k'} (u,v)$ is a binary mask that is unity at spatial frequencies where the image spectra, $G^{obs}_k(u,v)$ and $G^{obs}_{k'} (u,v)$, have an SNR above some threshold (we used SNR $>$ 5), and zero elsewhere. Similarly, $M^H_j (u,v)$ is a binary mask that is zero at spatial frequencies where the Fourier spectrum $G^{obs}_j (u,v)$ is zero, and is unity elsewhere.

During the minimization we further  control the number of variables by assuming that the amplitudes $A(u,v)$ are constant, and only solve for the variables $\phi(x,y)$ and $\psi(u,v)$.  We maintain this assumption through to the full MFBD step (see section \ref{sec:MFBD}).

\subsection{CMFBD: Step 2}
Once the object and PSFs for the control frames have been estimated, we can use the recovered object in a forward deconvolution problem (i.e., hold $F(u,v)$ fixed) to provide good initial PSF estimates for the non-control frames, $h_j(x,y)$. Similar to step 1, we enforce the prior knowledge that the PSFs should not depend on how they are estimated. We then minimize the error metric 
\begin{eqnarray}
\epsilon &=& \Sigma_j \Sigma_{x,y} \left[(g_j(x,y)-f(x,y) \odot h_j(x,y))/\sigma_k(x,y) \right]^2  \nonumber \\
&+& \gamma \Sigma_j\Sigma_{u,v} M_{k,j}(u,v) \nonumber \\
&|&P_j(u,v) \star P_j(u,v) - \Sigma_k H_k(u,v)\beta_{j,k}(u,v)/K|^2
\end{eqnarray}

\subsection{MFBD}
\label{sec:MFBD}
At the end of step 2, we have reasonable initial estimates for the object and all the PSFs. We then run a traditional MFBD with all the frames by minimizing equation (\ref{eqn:mfbd}).
This final step capitalizes on the information on the object that is available in the non-control frames. Because the MFBD algorithm is provided with high-quality estimates for the object and PSFs, the optimization algorithm has a better chance of not getting trapped in a local minimum far from the global minimum. We run the MFBD until we reach convergence, defined by a  change in error  metric of less than $10^{-5}$.  At this point, we restart the MFBD and allow $P(u,v)$ to vary without restricting the amplitudes. This is important for achieving the best restoration \citep{Jefferies2002}. Practically, we move from estimating the variables $\psi(u,v)$ to new variables $P_R(u,v)$ and $P_I(u,v)$, the real and imaginary parts of $P(u,v)$ \citep{Miura2003}.

We note that high-fidelity modeling of the speckle pattern requires that the restoration is performed on a pixel grid that samples the wavefront phase with at least 4-5 pixels sampling the atmospheric coherence length ($r_0$) for the observing conditions. It is also essential to accurately model the telescope aperture's details, including the structure associated with the spider supporting the secondary mirror.

\section{The Cloud Computational Architecture}
The workload of Kraken in terms of CPUs availability and type of access, which is mainly burst, is best suited in the context of Cloud Computing \citep{williams}. For this reason, following a similar approach as described in \citet{landoni18, landoni19}, we prepared a Platform as Service architecture on the Amazon Web Services (AWS)\footnote{\texttt{https://aws.amazon.com/}} cloud. In particular, we deployed a custom machine image with a fully working MATLAB license aiming to exploit, throughout the Parallel Computing Toolbox offered by the language, up to 96 vCPU cores available on the computer node on AWS. For the storage, we made use of both Elastic Block Storage (EBS) for the local storage of the node and Amazon Simple Storage Service (S3) for long-term archiving.

\section{Observations and results on $\alpha$ And binary}
For our first application of the MFBD to astronomical targets we chose the close binary star $\alpha$~And, which our team directly resolved through imaging for the first time in \citet{Mattioli_2019} by means of the Speckle Free Imaging technique (SFI) that we developed for processing SHARK-VIS image sequences \citep{Li_Causi_2017}. This binary is composed by a very bright $mag_R=2$ primary star, $\alpha$~And~A, and a $1.3\cdot10^{-1}$ contrast companion, $\alpha$~And~B, at a separation of $16.3 mas$ at the moment of the observations, comparable to the LBT theoretical diffraction limit of $\lambda/D=16.4 mas$ at the observation wavelength of $656nm$.

Observations were performed at LBT on 2018 December 23. A sequence of fast-cadence (1 kHz) short-exposure (1 ms) images were acquired continuously for 1 minute with the SHARK-VIS Forerunner instrument \citep{Pedichini_2017}, observing at $656nm$ with a $10nm$ bandwidth filter. A data set of 60,000 frames of 200x200 pixels size were aquired, corresponding to a field of view of $\sim 1~arcsec^2$, while the FLAO AO system \citep{Esposito_2010} was correcting 300 modes in a closed loop at 400 Hz. The seeing was unstable and above 1 arcsec, according to the DIMM telemetry.

Due to the poor seeing, a best-frames selection was necessary. To do this, we made a best-frame list, consisting of 3000 frames, by maximizing peak value and sharpness, and run the Kraken reconstruction of the first 219 of them, in order to be comparable with our previous result in \citet{Mattioli_2019}. We also did not perform PSF amplitude estimation during  the restoration process.

Thanks to the good quality of the acquired frames, the initial CMFBD steps only needed to use the best two frames out of these to provide excellent estimation of the object and the PSFs (see figure \ref{fig:kraken_vs_sfi}, panel (iv)). This allowed the complete Kraken MFBD to return the very clean image of the binary star shown in figure \ref{fig:kraken_vs_sfi}, panel (v), in which the two binary components are fully disentangled.
This result clearly surpasses our first imaging detection in \citet{Mattioli_2019} (shown in panel (iii) of the same figure for comparison), achieving an effective spatial resolution of $6.7 mas$ full width at half maximum, as measured from a two-gaussian fit of the reconstructed image.
It's also noticeable that no prior information on the nature of the source has been given to the algorithm, in which each pixel in the final image is estimated independently on the others.

\begin{figure}[]
    \centering
    \includegraphics[width=8cm]{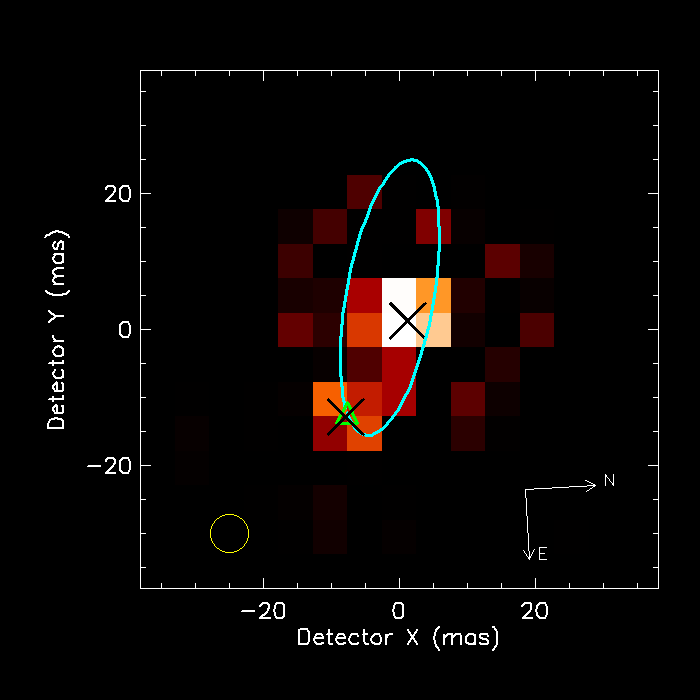}
    \caption{Spectro-interferometric orbit of $\alpha$~And~B overlaid on the Kraken MFBD reconstructed image, shown in log color scale. Black crosses indicate the gaussian-fitted positions for the two stars, while the green triangle is the expected location of the companion at the observation date. Sky orientation is shown by North and East arrows, while the circle in bottom left corner indicates the achieved FWHM resolution.}
    \label{fig:orbit}
\end{figure}


By fitting a double gaussian on this resulting image we derive a separation of $16.97 \pm 1.17 mas$ and a position angle $123 \pm 4.3\deg$. These numbers are compatible with the nominal expected position, derived from the orbital parameters given by \cite{Branham_2017} (position angle $126.7\deg$, separation $16.3 mas$), which were obtained by combining previous spectroscopic and interferometric observations (e.g. \citet{Pan_1992, Pourbaix_2004}), as shown in Figure \ref{fig:orbit}. Also, the flux ratio that we measured between the components, of $6.29 \pm 0.83$, is compatible with the expected value of $7.9 \pm 3.6$ (at $656nm$) considering the spectral types (B8IV and A3V) and ratio of radii ($1.64 \pm -0.23$) given in the literature \citep{Ryabchikova_1999}.

Errors in our photometric and astrometric result include both instrument uncertainties (namely a $\pm 1 \deg$ in the orientation of the Forerunner instrument and a $\pm 0.1 mas/pix$ in the estimated image scale of $5.08 mas/pix$), plus the variance of the MFBD image reconstruction, estimated by repeating the full Kraken restoration process on the next few subsets of 219 frames in our best-frames list.

\section{Conclusions}
\label{sec:Conclusions}
We presented the first application of the Kraken algorithm to the case of post adaptive optics high-resolution imaging in the visible on a large telescope, describing its mathematical basis and our implementation solutions.

Kraken is a high performace multi-frame blind deconvolution algorithm, working on fast-cadence frame sequences at KHz rate, which freeze the speckles evolution typical of the AO residual PSF.
It works by first computing a very precise initial guess by means of the Compact MFBD estimator, then following a complete MFBD procedure which performs a physical modeling of the instantaneous wavefront for each frame, ending up with an high-resolution reconstruction of the source brightness distribution.

We shown the result obtained on a fast-cadence observation of the spectroscopic binary $\alpha~And$ ($16.3 mas$ separation) acquired at the Large Binocular Telescope with the SHARK-VIS Forerunner instrument. Noticeably, the Kraken result clearly reconstructed two disentangled components without any prior information about the nature of the source, definitely surpassing the first direct detection that we obtained with the speckle-free imaging method. 

We demonstrated that photometry and astrometry performed on the Kraken reconstruction are compatible with those expected from the companion's orbit, already known by means of spectroscopy and interferometry. Also, the estimation of these important astronomical observables are very stable with respect to frames replacements. These facts together guarantee both the accuracy and the robustness of the Kraken output and its adequacy to be applied to high-resolution imaging in astronomy.

\acknowledgments

The Air Force Office of Scientific Research funded DAH and SMJ’s contributions to this work through contract FA9550-14-1-0178.
Cloud computing resources have been sustained by ICT-INAF (see \url{https://www.ict.inaf.it/computing/amazon-web-services-on-demand-access/}) .

%

\vspace{5mm}




\bibliography{Kraken}{}

\begin{thebibliography}{}
\expandafter\ifx\csname natexlab\endcsname\relax\def\natexlab#1{#1}\fi
\providecommand{\url}[1]{\href{#1}{#1}}
\providecommand{\dodoi}[1]{doi:~\href{http://doi.org/#1}{\nolinkurl{#1}}}
\providecommand{\doeprint}[1]{\href{http://ascl.net/#1}{\nolinkurl{http://ascl.net/#1}}}
\providecommand{\doarXiv}[1]{\href{https://arxiv.org/abs/#1}{\nolinkurl{https://arxiv.org/abs/#1}}}

\bibitem[{Branham(2016)}]{Branham_2017}
Branham, Richard~L., J. 2016, Monthly Notices of the Royal Astronomical
  Society, 464, 1095, \dodoi{10.1093/mnras/stw2393}

\bibitem[{Conan {et~al.}(2000)Conan, Fusco, Mugnier, \& Marchis}]{Conan:2000}
Conan, J.-M., Fusco, T., Mugnier, L., \& Marchis, F. 2000, The Messenger, 99,
  38

\bibitem[{{Esposito} {et~al.}(2010){Esposito}, {Riccardi}, {Fini}, {Puglisi},
  {Pinna}, {Xompero}, {Briguglio}, {Quir{\'o}s-Pacheco}, {Stefanini}, {Guerra},
  {Busoni}, {Tozzi}, {Pieralli}, {Agapito}, {Brusa-Zappellini}, {Demers},
  {Brynnel}, {Arcidiacono}, \& {Salinari}}]{Esposito_2010}
{Esposito}, S., {Riccardi}, A., {Fini}, L., {et~al.} 2010, in Society of
  Photo-Optical Instrumentation Engineers (SPIE) Conference Series, Vol. 7736,
  Adaptive Optics Systems II, ed. B.~L. {Ellerbroek}, M.~{Hart}, N.~{Hubin}, \&
  P.~L. {Wizinowich}, 773609, \dodoi{10.1117/12.858194}

\bibitem[{Hope \& Jefferies(2011)}]{Hope2011}
Hope, D.~A., \& Jefferies, S.~M. 2011, Opt. Lett., 36, 867,
  \dodoi{10.1364/OL.36.000867}

\bibitem[{Hope {et~al.}(2016)Hope, Jefferies, Hart, \& Nagy}]{Hope2016}
Hope, D.~A., Jefferies, S.~M., Hart, M., \& Nagy, J.~G. 2016, Opt. Express, 24,
  12116, \dodoi{10.1364/OE.24.012116}

\bibitem[{{Hope} {et~al.}(2019){Hope}, {Jefferies}, \& {Smith}}]{Hope2019}
{Hope}, D.~A., {Jefferies}, S.~M., \& {Smith}, C. 2019, Journal of the
  Astronautical Sciences, 66, 162, \dodoi{10.1007/s40295-018-00148-x}

\bibitem[{Ingleby \& McGaughey(2005)}]{Ingleby:05}
Ingleby, H.~R., \& McGaughey, D.~R. 2005, Opt. Lett., 30, 489,
  \dodoi{10.1364/OL.30.000489}

\bibitem[{{Jefferies} \& {Christou}(1993)}]{Jefferies1993}
{Jefferies}, S.~M., \& {Christou}, J.~C. 1993, \apj, 415, 862,
  \dodoi{10.1086/173208}

\bibitem[{Jefferies {et~al.}(2002)Jefferies, Lloyd-Hart, Hege, \&
  Georges}]{Jefferies2002}
Jefferies, S.~M., Lloyd-Hart, M., Hege, E.~K., \& Georges, J. 2002, Appl. Opt.,
  41, 2095, \dodoi{10.1364/AO.41.002095}

\bibitem[{Landoni {et~al.}(2018)Landoni, Genoni, Riva, Bianco, \&
  Corina}]{landoni18}
Landoni, M., Genoni, M., Riva, M., Bianco, A., \& Corina, A. 2018, in Software
  and Cyberinfrastructure for Astronomy V, ed. J.~C. Guzman \& J.~Ibsen, Vol.
  10707, International Society for Optics and Photonics (SPIE), 101 -- 109,
  \dodoi{10.1117/12.2312447}

\bibitem[{Landoni {et~al.}(2019)Landoni, Romano, Vercellone, Knödlseder,
  Bianco, Tavecchio, \& Corina}]{landoni19}
Landoni, M., Romano, P., Vercellone, S., {et~al.} 2019, The Astrophysical
  Journal Supplement Series, 240, 32, \dodoi{10.3847/1538-4365/aafcb5}

\bibitem[{{Li Causi} {et~al.}(2017){Li Causi}, Stangalini, Antoniucci,
  Pedichini, Mattioli, \& Testa}]{Li_Causi_2017}
{Li Causi}, G., Stangalini, M., Antoniucci, S., {et~al.} 2017, The
  Astrophysical Journal, 849, 85, \dodoi{10.3847/1538-4357/aa8e98}

\bibitem[{Matson(2008)}]{Matson2008}
Matson, C.~L. 2008, in Optics in Atmospheric Propagation and Adaptive Systems
  XI, ed. A.~Kohnle, K.~Stein, \& J.~D. Gonglewski, Vol. 7108, International
  Society for Optics and Photonics (SPIE), 178 -- 189,
  \dodoi{10.1117/12.801240}

\bibitem[{Mattioli {et~al.}(2018)Mattioli, Pedichini, Antoniucci, Causi,
  Piazzesi, Stangalini, \& Testa}]{Mattioli_2018}
Mattioli, M., Pedichini, F., Antoniucci, S., {et~al.} 2018, in Ground-based and
  Airborne Instrumentation for Astronomy VII, ed. C.~J. Evans, L.~Simard, \&
  H.~Takami, Vol. 10702, International Society for Optics and Photonics (SPIE),
  1332 -- 1347, \dodoi{10.1117/12.2312591}

\bibitem[{Mattioli {et~al.}(2019)Mattioli, Pedichini, Antoniucci, Causi,
  Piazzesi, Stangalini, Testa, Vaz, Pinna, Puglisi, Christou, \&
  Hinz}]{Mattioli_2019}
Mattioli, M., Pedichini, F., Antoniucci, S., {et~al.} 2019, Research Notes of
  the {AAS}, 3, 20, \dodoi{10.3847/2515-5172/ab0111}

\bibitem[{Miura(2003)}]{Miura2003}
Miura, N. 2003, Opt. Lett., 28, 2312, \dodoi{10.1364/OL.28.002312}

\bibitem[{{Pan} {et~al.}(1992){Pan}, {Shao}, {Colavita}, {Armstrong},
  {Mozurkewich}, {Vivekanand}, {Denison}, {Simon}, \& {Johnston}}]{Pan_1992}
{Pan}, X., {Shao}, M., {Colavita}, M.~M., {et~al.} 1992, \apj, 384, 624,
  \dodoi{10.1086/170904}

\bibitem[{Paxman {et~al.}(1992)Paxman, Schulz, \& Fienup}]{Paxman:92}
Paxman, R.~G., Schulz, T.~J., \& Fienup, J.~R. 1992, J. Opt. Soc. Am. A, 9,
  1072, \dodoi{10.1364/JOSAA.9.001072}

\bibitem[{{Pedichini} {et~al.}(2016){Pedichini}, {Ambrosino}, {Centrone},
  {Farinato}, {Li Causi}, {Pinna}, {Puglisi}, {Stangalini}, \&
  {Testa}}]{Pedichini_2016}
{Pedichini}, F., {Ambrosino}, F., {Centrone}, M., {et~al.} 2016, in Society of
  Photo-Optical Instrumentation Engineers (SPIE) Conference Series, Vol. 9908,
  Ground-based and Airborne Instrumentation for Astronomy VI, ed. C.~J.
  {Evans}, L.~{Simard}, \& H.~{Takami}, 990832, \dodoi{10.1117/12.2232375}

\bibitem[{{Pedichini} {et~al.}(2017){Pedichini}, {Stangalini}, {Ambrosino},
  {Puglisi}, {Pinna}, {Bailey}, {Carbonaro}, {Centrone}, {Christou},
  {Esposito}, {Farinato}, {Fiore}, {Giallongo}, {Hill}, {Hinz}, \&
  {Sabatini}}]{Pedichini_2017}
{Pedichini}, F., {Stangalini}, M., {Ambrosino}, F., {et~al.} 2017, \aj, 154,
  74, \dodoi{10.3847/1538-3881/aa7ff3}

\bibitem[{{Pinna} {et~al.}(2015){Pinna}, {Pedichini}, {Esposito}, {Centrone},
  {Puglisi}, {Farinato}, {Carbonaro}, {Agapito}, {Stangalini}, {Riccardi},
  {Xompero}, {Briguglio}, {Hinz}, {Bayley}, \& {Montoya}}]{Pinna_2015}
{Pinna}, E., {Pedichini}, F., {Esposito}, S., {et~al.} 2015, in Adaptive Optics
  for Extremely Large Telescopes IV (AO4ELT4), E58

\bibitem[{{Pourbaix, D.} {et~al.}(2004){Pourbaix, D.}, {Tokovinin, A. A.},
  {Batten, A. H.}, {Fekel, F. C.}, {Hartkopf, W. I.}, {Levato, H.}, {Morrell,
  N. I.}, {Torres, G.}, \& {Udry, S.}}]{Pourbaix_2004}
{Pourbaix, D.}, {Tokovinin, A. A.}, {Batten, A. H.}, {et~al.} 2004, A\&A, 424,
  727, \dodoi{10.1051/0004-6361:20041213}

\bibitem[{Ragazzoni(1996)}]{Ragazzoni_1996}
Ragazzoni, R. 1996, Journal of Modern Optics, 43, 289,
  \dodoi{10.1080/09500349608232742}

\bibitem[{Roggemann \& Welsh(1996)}]{Roggemann-1996}
Roggemann, M., \& Welsh, B. 1996, Imaging Through Turbulence (New York: CRC
  Press LLC)

\bibitem[{{Ryabchikova} {et~al.}(1999){Ryabchikova}, {Malanushenko}, \&
  {Adelman}}]{Ryabchikova_1999}
{Ryabchikova}, T.~A., {Malanushenko}, V.~P., \& {Adelman}, S.~J. 1999, \aap,
  351, 963

\bibitem[{Schulz(1993)}]{Schulz1993}
Schulz, T.~J. 1993, J. Opt. Soc. Am. A, 10, 1064,
  \dodoi{10.1364/JOSAA.10.001064}

\bibitem[{{Stangalini} {et~al.}(2016){Stangalini}, {Pedichini}, {Ambrosino},
  {Centrone}, \& {Del Moro}}]{Stangalini_2016}
{Stangalini}, M., {Pedichini}, F., {Ambrosino}, F., {Centrone}, M., \& {Del
  Moro}, D. 2016, in Society of Photo-Optical Instrumentation Engineers (SPIE)
  Conference Series, Vol. 9909, Adaptive Optics Systems V, ed. E.~{Marchetti},
  L.~M. {Close}, \& J.-P. {V{\'e}ran}, 99097P, \dodoi{10.1117/12.2232377}

\bibitem[{{van Noort} {et~al.}(2005){van Noort}, {Rouppe van der Voort}, \&
  {L{\"o}fdahl}}]{2005SoPh..228..191V}
{van Noort}, M., {Rouppe van der Voort}, L., \& {L{\"o}fdahl}, M.~G. 2005,
  \solphys, 228, 191, \dodoi{10.1007/s11207-005-5782-z}

\bibitem[{Williams {et~al.}(2018)Williams, Olsen, Khan, Pirone, \&
  Rosema}]{williams}
Williams, B.~F., Olsen, K., Khan, R., Pirone, D., \& Rosema, K. 2018, The
  Astrophysical Journal Supplement Series, 236, 4,
  \dodoi{10.3847/1538-4365/aab762}

\end{thebibliography}
\bibliographystyle{aasjournal}



\end{document}